\documentclass[fleqn,usenatbib]{mnras}

\usepackage{newtxtext,newtxmath}
\usepackage[T1]{fontenc}

\DeclareRobustCommand{\VAN}[3]{#2}
\let\VANthebibliography\thebibliography
\def\thebibliography{\DeclareRobustCommand{\VAN}[3]{##3}\VANthebibliography}

\usepackage{graphicx}	% Including figure files
\usepackage{amsmath}	% Advanced maths commands
\usepackage{siunitx}
\usepackage{comment}
\usepackage{color}

\newcommand{\unit}[1]{\,\rm{#1}}
\newcommand{\eq}[2][]{
	\begin{align#1}
		#2
	\end{align#1}
}

\newcommand{\myfig}[4][width=\columnwidth]{
    \begin{figure}
        \centering
            \includegraphics[#1]{#2}    % #2:画像の名前
        \caption{#3}                    % #3:キャプションの名前
        \label{#4} %ラベルをつけることにより参照できるようになります。
    \end{figure}
}

\newcommand{\diff}{\mathrm{d}}
\newcommand{\iu}{\mathrm{i}}
\newcommand{\e}{\mathrm{e}}

\newcommand{\cross}{\times}
\newcommand{\figref}[1]{Fig.~\ref{#1}}
\newcommand{\tabref}[1]{Table~\ref{#1}}
\newcommand{\secref}[1]{Sec.~\ref{#1}}

\definecolor{mygreen}{RGB}{0,130,0} 
\title[Early warning of precessing NSBH with 2.5G]{Early warning of precessing neutron-star black-hole binary mergers with the near-future gravitational-wave detectors}

\author[T. Tsutsui et al.]{
Takuya Tsutsui,$^{1,2}$\thanks{E-mail: tsutsui@resceu.s.u-tokyo.ac.jp}
Atsushi Nishizawa,$^{1}$
and Soichiro Morisaki$^{3}$
\\
$^{1}$Research Center for the Early Universe (RESCEU), Graduate School of Science, The University of Tokyo, Tokyo 113-0033, Japan\\
$^{2}$Department of Physics, Graduate School of Science, The University of Tokyo, Tokyo 113-0033, Japan\\
$^{3}$Department of Physics, University of Wisconsin-Milwaukee, Milwaukee, WI 53201, USA
}

\date{Accepted XXX. Received YYY; in original form ZZZ}

\pubyear{2021}

\begin{document}
\label{firstpage}
\pagerange{\pageref{firstpage}--\pageref{lastpage}}
\maketitle

% Abstract of the paper
\begin{abstract}
Since gravitational and electromagnetic waves from a compact binary coalescence carry independent information about the source, the joint observation is important for understanding the physical mechanisms of the emissions.
Rapid detection and source localization of a gravitational wave signal are crucial for the joint observation to be successful.
For a signal with a high signal-to-noise ratio, it is even possible to detect it before the merger, which is called early warning.
In this article, we estimate the performances of the early warning for neutron-star black-hole binaries, considering the precession effect of a binary orbit, with the near-future detectors such as A+, AdV+, KAGRA+, and Voyager.
We find that a gravitational wave source can be localized in $100 \unit{deg^2}$ on the sky before $\sim 10$--$40 \unit{s}$ of time to merger once per year.
\end{abstract}

% Select between one and six entries from the list of approved keywords.
% Don't make up new ones.

%%%%%%%%%%%%%%%%%%%%%%%%%%%%%%%%%%%%%%%%%%%%%%%%%%

%%%%%%%%%%%%%%%%% BODY OF PAPER %%%%%%%%%%%%%%%%%%

\section{Introduction} \label{SEC:introduction}
In 2017, two advanced LIGO~(aLIGO) detectors~\citep{LIGO1, LIGO2} and advanced Virgo~(AdV)~\citep{Virgo} detected a gravitational wave (GW) from a coalescence of binary neutron stars (BNS)~\citep{GW170817_observation}, dubbed as GW170817.
This event was followed up by many electromagnetic (EM) telescopes in broad bands and provided much information on an EM counterpart~\citep{GW170817_multimessenger}.
From the successful joint observation by GWs and EM waves, the association of a short gamma-ray burst (sGRB) with a merger of BNS was confirmed.
Furthermore, there are many benefits for not only sGRB but also the identification of the host galaxy, observations of kilonova~\citep{kilonovae}, measurement of Hubble constant~\citep{measure_hubble_constant} and so on.
It is desirable to increase the successful probability of the EM follow-up.

Rapid detection and source localization of a GW signal are crucial for the joint observation to be successful.
Then, one of the ways to improve the successful probability is early warning forecasting where and when a binary merger happens~\citep{early_warning, early_warning_simulation, early_warning_O4}.
With the alert, one can wait for the merger with EM telescopes pointed there.
Then we could observe prompt emission from a BNS merger~\citep{prompt_flash}, resonant shattering~\citep{resonant_shattering}, tidal disruption in neutron-star black-hole (NSBH) binaries~\citep{NSBH_EMemission}, and so on.

Now there are two aLIGO detectors, AdV, and KAGRA~\citep{KAGRA1}, called second-generation (2G) detectors.
The 2G detectors can detect GWs but not localize the sources before the merger well~\citep{early_warning_O4}.
Those detectors are planned to be upgraded to A+s~\citep{Aplus}, AdV+~\citep{Virgo_plus} and KAGRA+~\citep{KAGRA_plus} from the O5 observing run starting later than 2025.
Furthermore, Voyager~\citep{Voyager} is also planned the construction after 2030.
These GW detectors are called 2.5G detectors in this article.
They are more sensitive than the current detectors (\figref{FIG:PSD}) so that the early warning by them provides longer time to merger.
\myfig[width=0.9\linewidth]{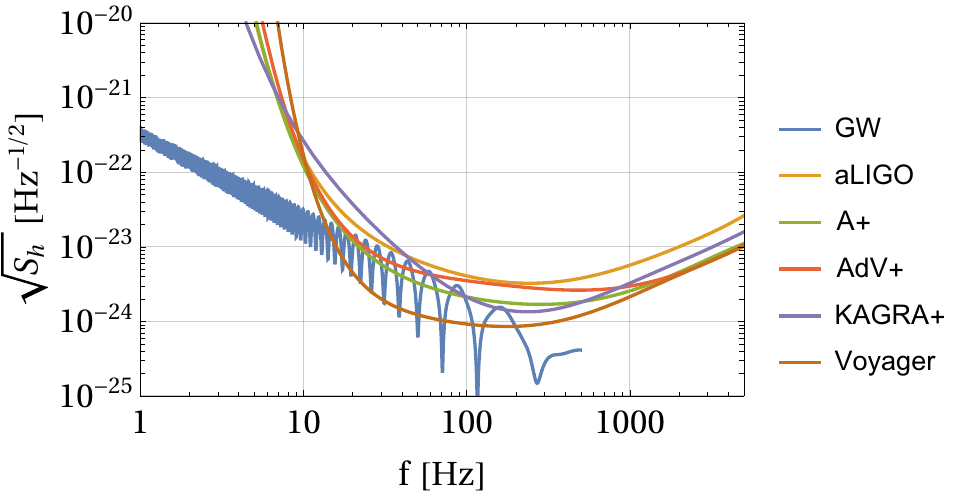}{Typical GW spectrum from a precessing NSBH merger at $z = 0.05$ (blue) and power spectral densities for aLIGO (orange), A+ (green), AdV+ (red), KAGRA+ (purple) and Voyager (brown)~\citep{LIGO1, Aplus, Virgo_plus, KAGRA_plus, Voyager}.}{FIG:PSD}
The early warning of BNS with the 2.5G detectors has been studied~\citep{early_warning_2.5G, early_warning_2.5G_dutycycle}.
Since the early warning is in the spotlight recently, there are many similar works: early detection with 2G detectors by neural networks ~\citep{early_warning_neural_network1, early_warning_neural_network2, early_warning_neural_network3}, estimations of the performances of early warning with third generation (3G) detectors~\citep{earlywarning_BNS, early_warning_precession, early_warning_3G_other_network} such as Einstein Telescope~\citep{ET_paper} and Cosmic Explorer~\citep{CE1}, early warning by considering higher modes of a GW~\citep{early_warning_higher_mode, early_warning_higher_mode_2G2.5G3G}.
Recently, NSBH and NSBH-like mergers have been detected~\citep{GW190814, NSBH_first_detection, GWTC3}, but their EM counterparts have not been reported yet.
Since the physics of the EM counterparts of the BNS and NSBH mergers could be different, we should consider the both possibilities as the targets of the early warning.

In this article, we consider the performance for NSBH binaries. For such asymmetric binaries, a precession effect can be important if either or both of compact objects have non-zero misaligned spins. Since the total angular momentum of a binary is not parallel with the orbital angular momentum, the orbital plane precesses and the amplitude of a GW from the binary is modulated~\citep{precession_KippThorne, precession1, precession2}.
By increasing information about the binary in the waveform, the estimation precision of binary parameters should be improved.

The organization of this article is as follows.
In \secref{SEC:analysis}, we review waveforms from precessing binaries and the Fisher analysis to estimate the performance of early warning to the waveforms.
We show in \secref{SEC:result} the parameter estimation errors as a function of frequency or time to merger for different networks of detectors and discuss the results in \secref{SEC:discussion}, compared with the current EM observations and previous studies on the early warning.
\secref{SEC:conclusion} is devoted to a summary.

\section{Analysis} \label{SEC:analysis}
Non-precessing binaries are frequently considered as GW sources for simplicity.
However, in reality, BHs may have non-zero spins.
Then a precessing spin induces the amplitude modulation of a GW and help improve the sky localization of a source.
We review the GW waveform from a precessing binary in \secref{SEC:waveform}.
The precession effect is strong for high mass-ratio binaries, and then we consider NSBH binaries.
In \secref{SEC:event_rate}, we estimate the number of NSBH binaries that we observe in a year and identify the highest-signal-to-noise-ratio (SNR) event in the realistic period of observations.
After that, we briefly explain the settings for estimating the performance of early warning with the Fisher matrix in \secref{SEC:fisher_analysis}.

\subsection{Waveform} \label{SEC:waveform}
An analysis in this article to estimate early warning performances is basically same as in~\citep{early_warning_precession}.
Thus, we briefly review the waveform in this subsection.
The readers can refer to~\citep{early_warning_precession} for the details.

For ease to read this article, we summarize the independent arguments set for the waveform:
\eq{
	\{ \lambda_i \} = \{ \mathcal{M}, \eta, \chi, t_c, \phi_c, d_\mathrm{L}, \theta, \phi, \psi, \theta_J, \kappa, \alpha_0 \} ,
}
$\mathcal{M}$ is the chirp mass, $\eta$ is the symmetric mass ratio, $\chi$ is the dimensionless spin magnitude of a heavier component (BH), $t_c$ is the coalescence time, $\phi_c$ is the coalescence phase, $d_\mathrm{L}$ is the luminosity distance, $\theta$ and $\phi$ are the longitude and latitude of GW source, $\psi$ is the polarization angle, $\theta_J$ is the inclination angle of total angular momentum, $\kappa$ is a inner product of unit vectors of the orbital and spin angular momentum and $\alpha_0$ is the initial precession angle.
We note that the spin magnitude of a lighter component (NS) is neglected here because it is much smaller than $\chi$ for NSBH binaries.
Also, the orbital eccentricity is not considered.

\begin{figure}
	\centering
	\includegraphics[width=0.8\linewidth]{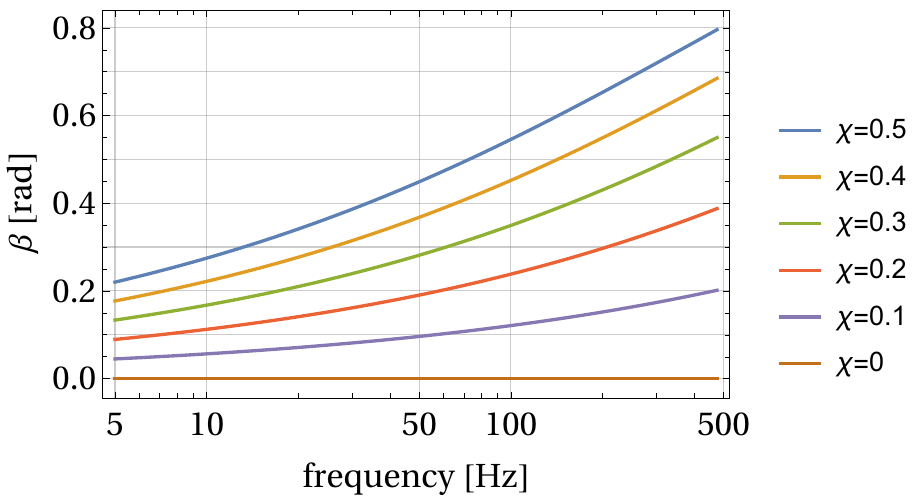} \\
	\includegraphics[width=0.8\linewidth]{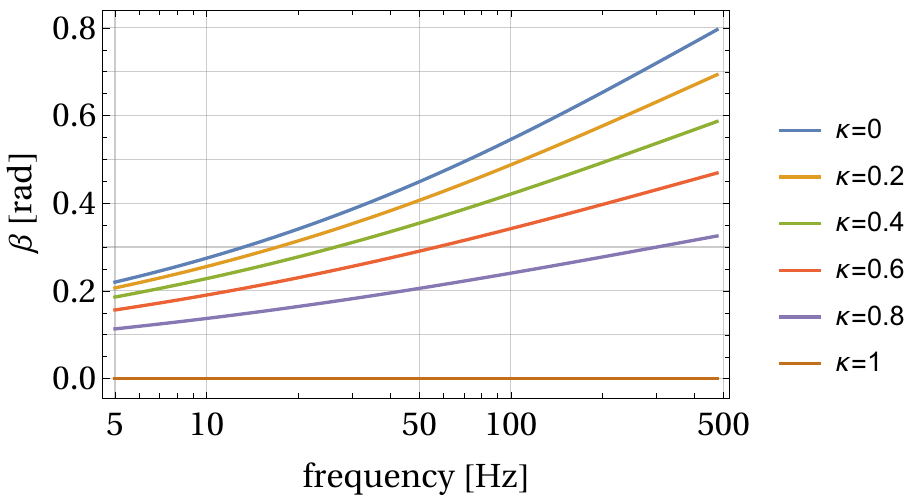}
	\caption{Precession angle $\beta$ as a function of frequency from $f_\mathrm{min}$ to $f_\mathrm{ISCO}$ for some $\chi$ (left) and $\kappa$ (right).}
	\label{FIG:result_f-beta}
\end{figure}

The GW waveform from a precessing compact binary is a product of the amplitude factor and the phase factor for non-precessing case and antenna response factor including the precession effect (see \figref{FIG:PSD} for the typical spectrum of a GW from the precessing binary):
\eq{
	\tilde{h}_I(f) =& Z_I(f) A(f)\, \e^{2\iu (\Phi_\mathrm{orb} - \zeta) + \phi_{I, \mathrm{Doppler}}} \, , \label{EQ:waveform} \\
	A(f) =& \left( \frac{5\pi}{24} \right)^{1/2} \frac{G^2 \mathcal{M}^2}{c^5 d_L} \left( \frac{\pi G\mathcal{M} f}{c^3} \right)^{-7/6} \, ,\\
	Z_I(f) =& \left( C_+(\alpha, \beta, \zeta) F_{I, +} + C_\cross(\alpha, \beta, \zeta) F_{I, \cross} \right) \nonumber \\
	& - \iu\left( S_+(\alpha, \beta, \zeta) F_{I, +} + S_\cross(\alpha, \beta, \zeta) F_{I, \cross} \right) \label{EQ:z}
}
The phase factor includes the orbital phase up to 1.5PN from~\citep{PNexample_1.5PN, early_warning_precession, PNexample_3.5PN_1, PNexample_3.5PN_2}, and the Doppler phase between the $I$-th detector at $\vec{r}_I$ and the geocenter, $\phi_{I, \mathrm{Doppler}} = -2\pi\iu f\vec{r}_I \cdot \vec{n} / c$.
The antenna response factor is written by mixtures of the antenna responses for non-precessing one, $F_{I,+/\cross}$, with coefficients originated from precession effect, $C_{+/\cross}$ and $S_{+/\cross}$~\citep{early_warning_precession, precession1, precession2, precession_KippThorne}.
Although there are three parameters ($\alpha, \beta, \zeta$) to express the precession effect, the one parameter is defined here because the other parameters are less important to estimate the performance of early warning~\citep{early_warning_precession}:
\eq{
    \beta(v) &= \cos^{-1} \left[ \frac{1+\kappa\gamma}{\Gamma_J} \right] \,, \label{EQ:beta}
}
where
\eq{
	\gamma(v) &= \frac{|\vec{S}|}{|\vec{L}|} = \frac{m_1 \chi}{m_2} v \, , \label{EQ:gamma}\\
	\Gamma_J(v) &= \frac{|\vec{J}|}{|\vec{L}|} = \sqrt{1+2\kappa\gamma+\gamma^2} \, ,\\
	\kappa &= \frac{\vec{L}}{|\vec{L}|} \cdot \frac{\vec{S}}{|\vec{S}|} \, ,
}
$v$ is the velocity of the component masses from Kepler law.
The frequency dependences of the $\beta$ for some parameters are in \figref{FIG:result_f-beta}.
Since $\chi = 0$ or $\kappa = 1$ case means non-precessing, $\beta$s are always zero.
We choose $0.5$ as our fiducial value of $\chi$ for our simulated NSBHs.
This value is consistent with the estimated primary spin of GW200115~\citep{NSBH_first_detection}.

\subsection{Event rate} \label{SEC:event_rate}
The number of sources available for the measurement of GW $N$ is estimated here as a function of redshift $z$.
The number of sources between redshifts $z$ and $z + \diff z$ is given by, e.~g.~\citep{dNdt_nishizawa},
\eq{
	\frac{\diff N}{\diff z} = \frac{4\pi c r(z)^2 \dot{n}(z)T_\mathrm{obs}}{(1 + z) H(z)} \label{EQ:dNdz}
}
where $T_\mathrm{obs}$ is the observation time, $\dot{n}(z)$ is the merger rate per unit comoving volume and unit proper time at redshift $z$, $r(z)$ is the comoving distance and $H(z)$ is the Hubble parameter.
% EM waves は全て見えるわけじゃないのでdetection efficiency $\epsilon(z)$ がついてたが、GW は全部見えるのでいらない（自明）
For the merger rate at $z < 1$, we take a linear evolution,
\eq{
	\dot{n}(z) = \dot{n}(0) (1 + 2z) \;,
}
which is based on the observation of the star formation history and is used in~\citep{dNdt_nishizawa}.
For NSBH binaries, we assume $\dot{n}(0) = 30 \unit{Gpc^{-3} yr^{-1}}$~\citep{rates_of_CBC} \footnote{Recently the merger rate from the observation of NSBH binaries was reported as $\dot{n}(0) = 45^{+75}_{-33}\unit{Gpc^{-3} yr^{-1}}$~\citep{NSBH_first_detection} when we were preparing the draft. Here we use the previous rate estimated from the population synthesis simulations, but our results hardly change if we take the median rate.}.
The integration of \eqref{EQ:dNdz} is shown in \figref{FIG:N_of_z}.
\myfig[width=0.9\linewidth]{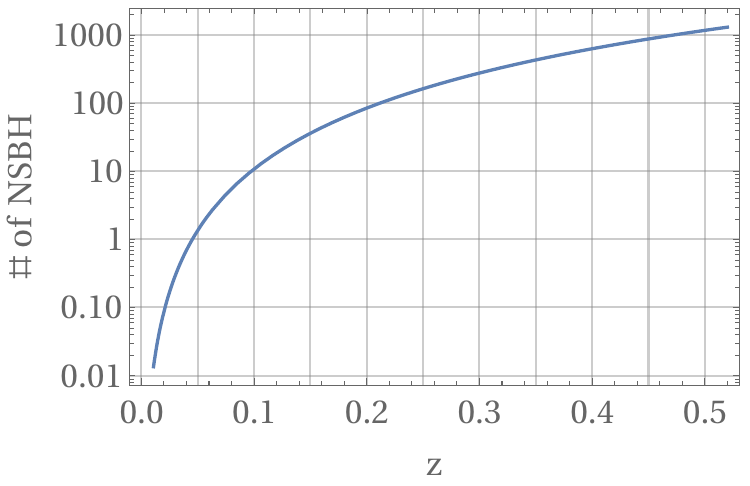}{The cumulative number of NSBH mergers per year as a function of redshift $z$.}{FIG:N_of_z}
A NSBH merger happens 1 per year at $z = 0.05$.
For comparison, we show $z_\mathrm{max}$ at which $\mathrm{SNR} = 8$ and the number of NSBH mergers for the detector networks in \tabref{TAB:zmax-number}.

\begin{table}
	\centering
	\caption{$z_\mathrm{max}$ and the average number of NSBH mergers per year for the different detector networks.  V+ means AdV+ and K+ means KAGRA+.}
	\begin{tabular}{c|cc} \hline \hline
		network & $z_\mathrm{max}$ & the average number of \\
		&& NSBH mergers [\unit{1/yr}] \\ \hline
		A+A+V+ & $0.185$ & $66.9$ \\
		A+A+V+K+ & $0.212$ & $99.9$ \\
		3 Voyagers & $0.517$ & $1274$ \\ \hline
	\end{tabular}
	\label{TAB:zmax-number}
\end{table}

\subsection{Fisher analysis} \label{SEC:fisher_analysis}
To estimate the parameter errors, we use the Fisher information matrix, $F_{ij}$, defined between $f_\mathrm{min} = 5 \unit{Hz}$ and the innermost stable circular orbit frequency for a Schwarzschild BH, $f_\mathrm{ISCO} = \frac{1}{12\sqrt{6}\pi} \frac{c^3}{G(m_1 + m_2)}$:
\eq{
	F_{ij}(f) &= 4 \Re \int_{f_\mathrm{min}}^{f} \frac{\partial_{i}\tilde{h}^*(f') \partial_{j}\tilde{h}(f')}{S_n(f')} \,\diff f' + \frac{\delta_{ij}}{(\delta \lambda_i)^2} \;,
}
where $\partial_{i}$ is a partial derivative with respect to the parameter $\lambda_i$.
Although $f_\mathrm{ISCO}$ ideally depends on $\chi$~\citep{ISCO_frequency_KerrBH}, the dependence can be neglected because our interest is early warning, that is, the low-frequency region of the Fisher matrix $F_{ij}$.
We have taken into account the prior effects (the second term) as done in~\citep{early_warning_precession} where $S_n(f)$ is a power spectral density, $\delta_{ij}$ is the Kronecker delta, and $\delta\lambda_i$ is a $1\sigma$ error of a Gaussian distribution and is taken as a physical range of the parameter.
From the Cramer-Rao inequality~\citep{fisher_CBC}, the inverse Fisher matrix $F_{ij}^{-1}$ can be recognized as the covariance matrix.

In this article, we quantify the performance of early warning by the 1-sigma statistical errors of the luminosity distance, the sky localization error, $\Delta d_\mathrm{L}$ and $\Delta\Omega$, and the number of galaxies in the error volume, $\Delta N$, as functions of time to merger 
\eq{
	t_c - t = \frac{5}{256} (\pi f)^{-8/3} \left( \frac{G \mathcal{M}}{c^3} \right)^{-5/3}\;, \label{EQ:time_to_merger}
}
at the Newtonian order~\citep{maggiore1, Jolien}.
The Fisher matrix is calculated for $200$ samples\footnote{The choice of the number of samples is due to computation time. As the precession effect is not simple, we take $\sim 2$ days for one network and one parameter set with a reasonable computer. So we have to spend O(a month) in total. Since the estimations are from $200$ samples, the results provided in this article have $\sim 7\%$ error.} with the following choice of fiducial parameters of a spinning BH and a non-spinning NS: the angles, $(\theta, \phi)$ and $(\theta_J, \psi)$, are isotropically sampled.
The masses and spins are $\chi = 0.5$, $\kappa = 0$, $m_1 = 7 \unit{M_\odot}$ and $m_2 = 1.4 \unit{M_\odot}$. 
The redshift $z$ is $0.05$ from the discussions in \secref{SEC:event_rate}.
In \secref{SEC:other_parameters}, we also consider the performance for other parameters, $\chi = 0.1, 0.2$ and $\kappa = \cos 30^\circ, \cos 45^\circ$, and verify that the estimation errors are not degraded significantly.
We consider the 2.5G detectors: A+s, AdV+, KAGRA+, and Voyager at Hanford, Livingston and Virgo sites. They are the upgrades of the current detectors at the same sites.
Also, as with~\citep{early_warning_precession}, the Earth rotation effect is taken into account, although the observable duration is much shorter than the previous case and the effect is almost negligible.

\section{Result} \label{SEC:result}
\subsection{network dependence} \label{SEC:network_dependence}
\begin{figure}
	\centering
	\begin{tabular}{c}
	\begin{minipage}{0.8\linewidth}
		\includegraphics[width=\linewidth]{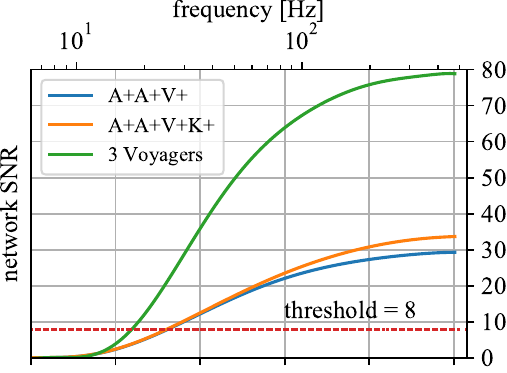} \\
		\includegraphics[width=\linewidth]{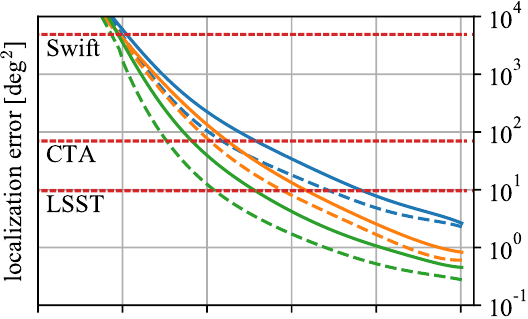} \\
		\includegraphics[width=\linewidth]{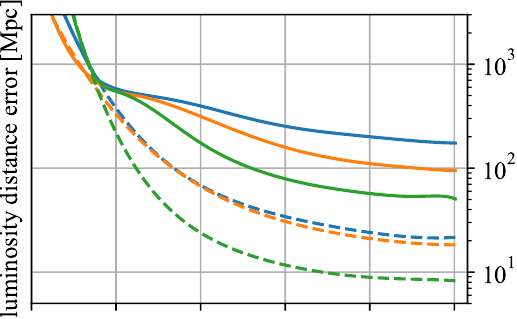} \\
		\includegraphics[width=\linewidth]{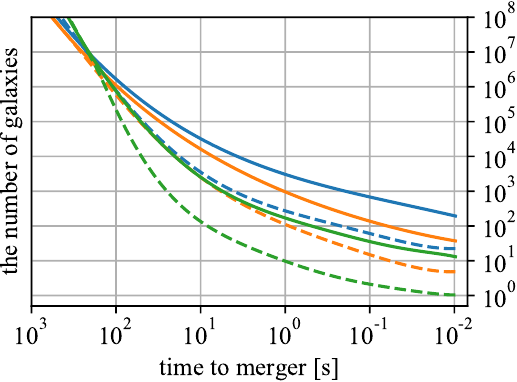}
	\end{minipage}
	\end{tabular}
	\caption{Medians of localization errors with A+s at Hanford and Livingston sites and AdV+ (blue), those plus KAGRA+ (orange), and three Voyagers at Hanford, Livingston, and Virgo sites (green) as a function of time to merger. The panels from the top to the bottom are for the network SNR without the precession effect, the sky localization error, the distance error, and the number of galaxies in the error volume. Solid (dashed) lines are without (with) the precession effect. The red dotted line in the network SNR plot represents the SNR threshold for detection at $8$. The red dotted lines in the sky localization plot represent the typical field of views of Swift, CTA, and LSST ($4900 \unit{deg^2}$, $70 \unit{deg^2}$, and $9.6 \unit{deg^2}$), respectively.}
	\label{FIG:result_TimeEvolution}
\end{figure}

\begin{table}
	\centering
	\caption{Medians of time-to-merger for NSBH binaries when the conditions for the network SNR, $\Delta\Omega$, and $\Delta N$ are satisfied.  The values in this table are in the unit of seconds.  ``w/o'' and ``w/'' mean ``without'' and ``with'' the precession effect, respectively.}
	\begin{tabular}{l|cc|cc|cc} \hline \hline
		& \multicolumn{6}{c}{time to merger [s]} \\ \hline
		network & \multicolumn{2}{c|}{A+A+V+} & \multicolumn{2}{c|}{A+A+V+K+} & \multicolumn{2}{c}{3 Voyager} \\ \hline
		precession & w/o & w/ & w/o & w/ & w/o & w/ \\ \hline
		network SNR$=8$                   & $24$   & $24$   & $25$   & $25$   & $64$   & $64$ \\ \hline
		$\Delta\Omega = 300 \unit{deg^2}$ & $13$   & $23$   & $19$   & $25$   & $35$   & $57$ \\
		$\Delta\Omega = 100 \unit{deg^2}$ & $4.2$  & $9.6$  & $7.9$  & $12$   & $19$   & $36$ \\ \hline
		$\Delta N = 10^4$                 & $3.7$  & $18$   & $7.2$  & $20$   & $20$   & $50$ \\
		$\Delta N = 10^3$                 & $0.20$ & $4.1$  & $1.1$  & $5.8$  & $5.5$  & $25$ \\
		$\Delta N = 10^2$                 & -      & $0.22$ & $0.07$ & $0.90$ & $0.49$ & $8.3$ \\ \hline
	\end{tabular}
	\label{TAB:NSBH_golden}
\end{table}

Figure~\ref{FIG:result_TimeEvolution} are the medians of the network SNR, the sky localization errors $\Delta\Omega$ and distance errors $\Delta d_\mathrm{L}$, and the number of galaxies in the error volume $\Delta N$, for the detector networks, (A+, A+, AdV+), (A+, A+, AdV+, KAGRA+), and 3 Voyagers.
Since the precession effect affects the variance of SNR but not the median (the measured relative difference is $\lesssim 2\%$), the network SNRs are plotted only for non-precessing cases in \figref{FIG:result_TimeEvolution}.
To convert the error volume to the number of galaxies $\Delta N$, the number density of the galaxies is assumed to be $0.01 \unit{Mpc^{-3}}$ as explained in more detail in~\citep{Measurement_HubbleConstant}.
The time to mergers when $\Delta\Omega$ and $\Delta N$ reach typical values are summarized in \tabref{TAB:NSBH_golden}.

GW detection alerts are generated when the network SNR exceeds $8$ and can be sent out before the merger at $24 \unit{s}$ for two A+s and AdV+, $25 \unit{s}$ for two A+s, AdV+ and KAGRA+, and $64 \unit{s}$ for three Voyagers.
At that times, for a precessing (non-precessing) case, the sky localization errors are $323 \unit{deg^2}$($625 \unit{deg^2}$), $300 \unit{deg^2}$($464 \unit{deg^2}$), and $423 \unit{deg^2}$($1158 \unit{deg^2}$), respectively.

In \figref{FIG:result_TimeEvolution}, the estimations with the precession effect are much improved, especially for $\Delta d_\mathrm{L}$ then $\Delta N$.
However, the improvement for $\Delta\Omega$ is relatively smaller because $\Delta\Omega$ has already been well determined by arrival-timing measurements of a GW, and the precession effect does not help much.
On the other hand, the degeneracy between the luminosity distance and the inclination angle is broken by precession~\citep{precession_breaks_degeneracies_LISA, precession_breaks_degeneracies_ground}, and then $\Delta d_\mathrm{L}$ is significantly improved.

\begin{figure}
	\centering
	\includegraphics[width=0.8\linewidth]{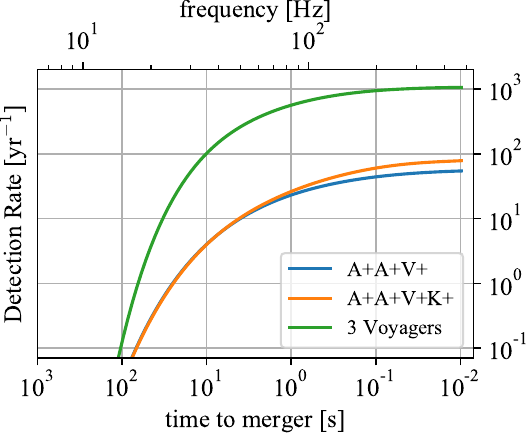}
	\caption{Detection rate for non-precessing binaries with A+s at Hanford and Livingston sites and AdV+ (blue), those plus KAGRA+ (orange), and three Voyagers at Hanford, Livingston, and Virgo sites (green) as a function of time to merger.  Explanations for the width of distributions are written in \secref{SEC:network_dependence}}
	\label{FIG:result_DetectionRate}
\end{figure}
Figure~\ref{FIG:result_DetectionRate} is the detection rate for the detector networks, that is, the expected number of the GW sources with the SNR higher than the detection threshold $8$.
As in the plot of network SNR in \figref{FIG:result_TimeEvolution}, the detection rates are for only a non-precessing case.
From \figref{FIG:result_DetectionRate}, we can find how many events are detectable at a time-to-merger, e.g. the GW events with the detection rate $10\unit{yr^{-1}}$, which corresponds to $z \sim 0.1$ from \figref{FIG:N_of_z}, can be detected at \SI{3.0}{s} for A+s and AdV+, \SI{3.7}{s} for those plus KAGRA+ and \SI{30}{s} for three Voyagers, before the merger.
From the uncertainties of SNRs in $200$ samples, the lines in \figref{FIG:result_DetectionRate} have the width of distributions ($1\sigma$) for the precessing (non-precessing) case; $68\%$ ($84\%$) for A+s and AdV+, $53\%$ ($73\%$) for those plus KAGRA+ and $58\%$ ($78\%$) for three Voyagers.

\subsection{parameter dependence} \label{SEC:other_parameters}
\begin{figure}
	\centering
	\includegraphics[width=0.8\linewidth]{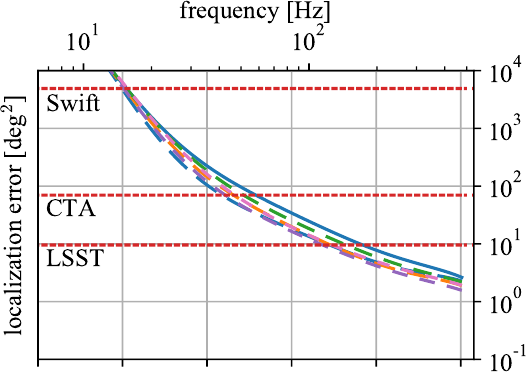} \\
	\includegraphics[width=0.8\linewidth]{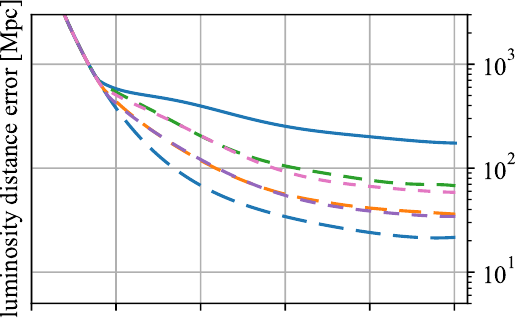} \\
	\includegraphics[width=0.8\linewidth]{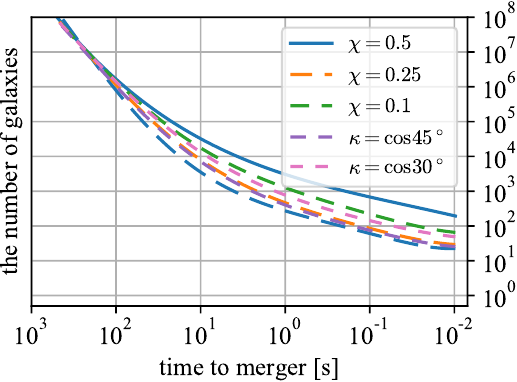}
	\caption{Medians of localization errors with A+s at Hanford and Livingston sites and AdV+ for some parameters.  The solid line is for non-precessing case and the dashed lines are for precessing case.  The blues are for $(\chi, \kappa) = (0.5, 0)$, the orange for $(0.25, 0)$, the green for $(0.1, 0)$, the purple for $(0.5, \cos45^\circ)$, pink for $(0.5, \cos30^\circ)$}
	\label{FIG:result_TimeEvolution_otherparams_ApApVp}
\end{figure}
\begin{figure}
	\centering
	\includegraphics[width=0.8\linewidth]{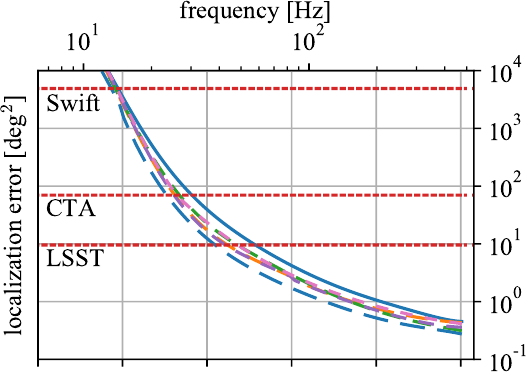} \\
	\includegraphics[width=0.8\linewidth]{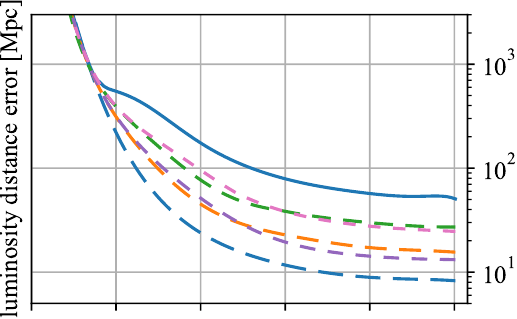} \\
	\includegraphics[width=0.8\linewidth]{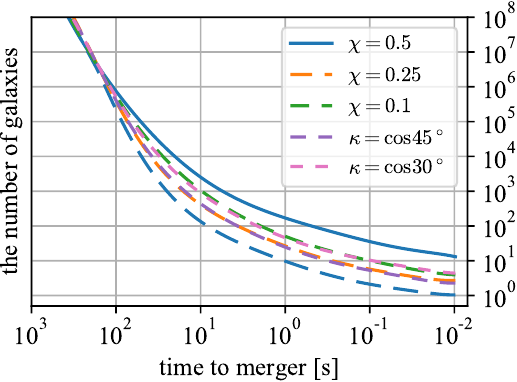}
	\caption{Medians of localization errors with three Voyagers at Hanford, Livingston and Virgo sites for some parameters. The lines and parameters chosen are the same as those in Fig.~\ref{FIG:result_TimeEvolution_otherparams_ApApVp}.}
	\label{FIG:result_TimeEvolution_otherparams_3Voyager}
\end{figure}

In the previous subsection, we considered our fiducial set of source parameters.
However, in reality, mass and spin parameters have the distributions depending on the astrophysical formation channel and are still highly uncertain~\citep{NSBH_first_detection}.
To show parameter dependences, we estimate the performances for less-precessing NSBHs with some $\chi$ and $\kappa$.

$\chi$ does not affect the SNR evolutions in our approximated waveform because it does not appear in GW amplitude.
Also, $\kappa$ does not affect the median of the SNR evolution although the variance of SNR is affected (same as \secref{SEC:network_dependence}).
Then, the SNR evolutions are same as those in Fig.~\ref{FIG:result_TimeEvolution}.
The estimated performances of $\Delta\Omega$, $\Delta d_\mathrm{L}$, and $\Delta N$ are in \figref{FIG:result_TimeEvolution_otherparams_ApApVp} and \ref{FIG:result_TimeEvolution_otherparams_3Voyager}.
For simplicity, the performances for non-precessing case except for $\chi = 0.5$ and $\kappa = 0$ are omitted in both figures.
In both figures, the improvements of $\varDelta\Omega$ are almost same because the maximum improvement in the case of $\chi = 0.5$ is not large, that is, the performances for $\varDelta\Omega$ are robust even for less-precessing parameters.

For $\Delta d_\mathrm{L}$ and $\Delta N$, the result for $(\chi, \kappa) = (0.25, \cos90^\circ)$ and $(0.5, \cos30^\circ)$ are almost same with that for $(0.5, \cos45^\circ)$ and $(0.1, \cos90^\circ)$, respectively.
This is because the time evolutions of $\beta$ for those are almost same between those parameter sets (see \figref{FIG:result_f-beta}).
In other word, the improvements for $\Delta d_\mathrm{L}$ and $\Delta N$ are proportional to $\beta$, so that we can obtain performances for other parameters from Eq.~\eqref{EQ:beta}.

Also the component masses can be changed.
However, the chirp mass is varied by $\sim 30\%$ in the mass range that EM radiation is expected~\citep{remnant_mass_function}.
Then the difference of the performances for the range is $\sim 30\%$~\citep{early_warning_precession}.

\section{Discussions} \label{SEC:discussion}
\subsection{EM telescopes}
Here we compare our results of the sky localization with the field of view (FoV) of typical EM telescopes.   
The FoV is $\sim 70 \unit{deg^2}$ for CTA-SST~\citep{CTA, CTA_FOV}, $\sim 4900 \unit{deg^2}$ for Swift-BAT~\citep{Swift_BAT}, $\sim \SI{2.6e4}{deg^2}$ for Fermi-GBM~\citep{Fermi_GBM}, and $\sim 9.6 \unit{deg^2}$ for LSST~\citep{LSST}.
Swift and Fermi are all-sky sweeping surveys.
CTA can point to the EM counterpart in $\lesssim 1 \unit{min}$.
LSST can tilt by $3.5 \unit{deg}$ in $5 \unit{s}$ and needs $20 \unit{s}$ -- $40 \unit{s}$ for the exposure. From \figref{FIG:result_TimeEvolution}, for the all detector networks, the EM counterpart can be inside the FoV of Swift and Fermi before $\sim 100 \unit{s}$ of the merger.
On the other hand, it can be inside that of CTA and LSST before $\mathcal{O}(0.1 - 1) \unit{s}$ for a non-precessing case and $\mathcal{O}(1 - 10) \unit{s}$ for a precessing case.
Thus, with the early warning for NSBH binary mergers, precursor emissions can be observed before the merger by Swift and Fermi, and prompt emissions and afterglows from sGRB can be observed by CTA and LSST.

So far we have not taken into account latencies produced by the data acquisition of GW detectors and EM telescopes, the detection of a GW signal, and the localization of a GW source.
Especially, the parameter inference taking into account the precession effects is computationally costly, and takes around half a day even with a fast method to evaluate likelihood \citep{localization_precession}.
The development of a fast localization technique incorporating the precession effect is necessary for early warnings of NSBH events.

\subsection{GW Detector networks} \label{SEC:discussion_4th_detector}
In \figref{FIG:result_TimeEvolution}, the results for two A+s and AdV+ are similar to those for two A+s, AdV+, and KAGRA+.
However, it does not mean that a fourth detector is unnecessary for multi-messenger observations.
The first advantage is duty cycle.
The duty cycle that at least three active detectors out of four detectors is much higher than the one that three active detectors out of three detectors, i.~e.~$51\%$ increases to $82\%$ given a single-detector duty cycle of $80\%$.
The second advantage is multimodality in the sky localization.
When three detectors are at work and only the arrival-time differences between detectors are measured, a mirror candidate spot in the sky probability map always appears above/below the plane spanned by the detectors.
By adding a fourth detector to the network, a mirror candidate spot can be eliminated. Then EM observers can avoid pointing their telescopes to the fake sky direction.
From these two reasons, the fourth detector is important for the multi-messenger astronomy.

\subsection{Comparison with previous studies} \label{SEC:discussion_comparison}
The similar study on the early warning with the 2.5G detector networks but for BNS has been done in \citep{early_warning_2.5G}.
The times of the early warning before merger for $\Delta\Omega = 100 \unit{deg^2}$ are $54 \unit{s}$ for two A+s and AdV (not AdV+) and $194 \unit{s}$ for three Voyagers.
On the other hand, our results for $\Delta\Omega = 100 \unit{deg^2}$ without the precession effect are $4.2 \unit{s}$ for two A+s and AdV+ and $19 \unit{s}$ for three Voyagers.
Their results are much better than ours.
The most dominant reason for the difference is the sources considered. We assume NSBH binaries but they assume BNS. Since the chirp mass of the latter is smaller and its frequency evolution is slower, BNS provide a factor of $4$ longer time-to-merger than NSBH binaries, as explicitly seen in~\eqref{EQ:time_to_merger}.
There are minor differences which can explain another factor of the difference: the source distance, the credible levels of sky localization error, the sensitivity of Voyager, the detector networks.
Since an EM counterpart of NSBH mergers is different from that of the BNS merger, we should consider them as targets for the early warning.

For 3G detectors that will start observations from the mid in 2030s, the early warning for NSBH binary coalescences at $z=0.1$ can be done typically with the times to merger of $12$ -- $15 \unit{minutes}$ and $50$ -- $300 \unit{seconds}$ when the sky localization areas reach $100\unit{deg}^2$ and $10 \unit{deg^2}$, respectively~\citep{early_warning_precession}.
In addition, not only $\Delta d_\mathrm{L}$ but also $\Delta\Omega$ are significantly improved due to the precessing effect.
This is because the sensitive frequency band of the 3G detectors is much broader than that of the 2.5G detectors, particularly at lower frequencies, and allows us to observe NSBH binaries for $\sim 8 \unit{hours}$. Then the Earth rotation and precession effects can break parameter degeneracies much more efficiently.

\section{Conclusion} \label{SEC:conclusion}
The combination of GW and EM observations can provide much information about the source.
Rapid detection and source localization of a GW signal are crucial for the joint observation to be successful.
It can be done by early warning which forecasts where and when a binary merger happens before the merger~\citep{early_warning, early_warning_simulation, early_warning_O4}.
We estimate the performances of the early warning for some detector networks from A+s, AdV+, KAGRA+ and Voyager for precessing and non-precessing NSBH binaries. As a result, we found that the direction can be known with $\Delta\Omega = 100 \unit{deg^2}$ in $\sim 10 \unit{s}$--$40 \unit{s}$ before the merger.
The improvement by considering the precession effect is about a factor $\sim 2$ for the sky localization error but $\sim 10$ for the distance error.
Therefore the number of galaxies in the error volume is much reduced by considering the precession.

\section*{Acknowledgments}
T.~T. is supported by International Graduate Program for Excellence in Earth-Space Science (IGPEES).
A.~N. is supported by JSPS KAKENHI Grant Nos. JP19H01894 and JP20H04726 and by Research Grants from Inamori Foundation.
S.~M. is supported by NSF PHY-1912649.

\section*{Data availability}
The data underlying this article will be shared on reasonable request to the corresponding author.

%%%%%%%%%%%%%%%%%%%% REFERENCES %%%%%%%%%%%%%%%%%%

% The best way to enter references is to use BibTeX:

\bibliographystyle{mnras}
\bibliography{references} % if your bibtex file is called example.bib

\begin{thebibliography}{}
\makeatletter
\relax
\def\mn@urlcharsother{\let\do\@makeother \do\$\do\&\do\#\do\^\do\_\do\%\do\~}
\def\mn@doi{\begingroup\mn@urlcharsother \@ifnextchar [ {\mn@doi@}
  {\mn@doi@[]}}
\def\mn@doi@[#1]#2{\def\@tempa{#1}\ifx\@tempa\@empty \href
  {http://dx.doi.org/#2} {doi:#2}\else \href {http://dx.doi.org/#2} {#1}\fi
  \endgroup}
\def\mn@eprint#1#2{\mn@eprint@#1:#2::\@nil}
\def\mn@eprint@arXiv#1{\href {http://arxiv.org/abs/#1} {{\tt arXiv:#1}}}
\def\mn@eprint@dblp#1{\href {http://dblp.uni-trier.de/rec/bibtex/#1.xml}
  {dblp:#1}}
\def\mn@eprint@#1:#2:#3:#4\@nil{\def\@tempa {#1}\def\@tempb {#2}\def\@tempc
  {#3}\ifx \@tempc \@empty \let \@tempc \@tempb \let \@tempb \@tempa \fi \ifx
  \@tempb \@empty \def\@tempb {arXiv}\fi \@ifundefined
  {mn@eprint@\@tempb}{\@tempb:\@tempc}{\expandafter \expandafter \csname
  mn@eprint@\@tempb\endcsname \expandafter{\@tempc}}}

\bibitem[\protect\citeauthoryear{Aasi et~al.,}{Aasi et~al.}{2015}]{LIGO1}
Aasi J.,  et~al., 2015, \mn@doi [Classical and Quantum Gravity]
  {10.1088/0264-9381/32/7/074001}, 32, 074001

\bibitem[\protect\citeauthoryear{Abadie, Abbott  et~al.}{Abadie
  et~al.}{2010}]{rates_of_CBC}
Abadie J.,  Abbott B.~P.,   et~al., 2010, \mn@doi [Classical and Quantum
  Gravity] {10.1088/0264-9381/27/17/173001}, 27, 173001

\bibitem[\protect\citeauthoryear{Abbott et~al.,}{Abbott et~al.}{2016}]{Aplus}
Abbott B.~P.,  et~al., 2016, \mn@doi [Living Reviews in Relativity]
  {10.1007/lrr-2016-1}, 19

\bibitem[\protect\citeauthoryear{Abbott et~al.}{Abbott et~al.}{2017a}]{CE1}
Abbott B.~P.,  et~al., 2017a, \mn@doi [Classical and Quantum Gravity]
  {10.1088/1361-6382/aa51f4}, 34, 044001

\bibitem[\protect\citeauthoryear{Abbott et~al.,}{Abbott
  et~al.}{2017b}]{GW170817_observation}
Abbott B.,  et~al., 2017b, \mn@doi [Physical Review Letters]
  {10.1103/physrevlett.119.161101}, 119

\bibitem[\protect\citeauthoryear{Abbott et~al.}{Abbott
  et~al.}{2017c}]{GW170817_multimessenger}
Abbott B.~P.,  et~al., 2017c, \mn@doi [Astrophys. J.]
  {10.3847/2041-8213/aa91c9}, 848, L12

\bibitem[\protect\citeauthoryear{Abbott, Abbott  et~al.}{Abbott
  et~al.}{2020}]{GW190814}
Abbott R.,  Abbott T.~D.,   et~al., 2020, \mn@doi [The Astrophysical Journal]
  {10.3847/2041-8213/ab960f}, 896, L44

\bibitem[\protect\citeauthoryear{Abbott, Abbott  et~al.}{Abbott
  et~al.}{2021a}]{GWTC3}
Abbott R.,  Abbott T.~D.,   et~al., 2021a, {GWTC-3: Compact Binary Coalescences
  Observed by LIGO and Virgo During the Second Part of the Third Observing Run}
  (\mn@eprint {arXiv} {2111.03606})

\bibitem[\protect\citeauthoryear{Abbott, Abbott  et~al.}{Abbott
  et~al.}{2021b}]{NSBH_first_detection}
Abbott R.,  Abbott T.~D.,   et~al., 2021b, \mn@doi [The Astrophysical Journal
  Letters] {10.3847/2041-8213/ac082e}, 915, L5

\bibitem[\protect\citeauthoryear{Acernese et~al.,}{Acernese
  et~al.}{2015}]{Virgo}
Acernese F.,  et~al., 2015, \mn@doi [Classical and Quantum Gravity]
  {10.1088/0264-9381/32/2/024001}, 32, 024001

\bibitem[\protect\citeauthoryear{Acharyya, Agudo  et~al.}{Acharyya
  et~al.}{2019}]{CTA_FOV}
Acharyya A.,  Agudo I.,   et~al., 2019, \mn@doi [Astroparticle Physics]
  {https://doi.org/10.1016/j.astropartphys.2019.04.001}, 111, 35

\bibitem[\protect\citeauthoryear{Adhikari et~al.,}{Adhikari
  et~al.}{2020}]{Voyager}
Adhikari R.~X.,  et~al., 2020, \mn@doi [Classical and Quantum Gravity]
  {10.1088/1361-6382/ab9143}, 37, 165003

\bibitem[\protect\citeauthoryear{Apostolatos, Cutler, Sussman  \&
  Thorne}{Apostolatos et~al.}{1994}]{precession_KippThorne}
Apostolatos T.~A.,  Cutler C.,  Sussman G.~J.,   Thorne K.~S.,  1994, \mn@doi
  [Phys. Rev. D] {10.1103/PhysRevD.49.6274}, 49, 6274

\bibitem[\protect\citeauthoryear{Baltus, Janquart, Lopez, Reza, Caudill  \&
  Cudell}{Baltus et~al.}{2021}]{early_warning_neural_network2}
Baltus G.,  Janquart J.,  Lopez M.,  Reza A.,  Caudill S.,   Cudell J.-R.,
  2021, \mn@doi [Phys. Rev. D] {10.1103/PhysRevD.103.102003}, 103, 102003

\bibitem[\protect\citeauthoryear{Barthelmy et~al.,}{Barthelmy
  et~al.}{2005}]{Swift_BAT}
Barthelmy S.~D.,  et~al., 2005, \mn@doi [Space Science Reviews]
  {10.1007/s11214-005-5096-3}, 120, 143–164

\bibitem[\protect\citeauthoryear{Bartos et~al.,}{Bartos et~al.}{2018}]{CTA}
Bartos I.,  et~al., 2018, \mn@doi [Monthly Notices of the Royal Astronomical
  Society] {10.1093/mnras/sty602}, 477, 639–647

\bibitem[\protect\citeauthoryear{Brown, Lundgren  \& O'Shaughnessy}{Brown
  et~al.}{2012}]{precession2}
Brown D.~A.,  Lundgren A.,   O'Shaughnessy R.,  2012, \mn@doi [Phys. Rev. D]
  {10.1103/PhysRevD.86.064020}, 86, 064020

\bibitem[\protect\citeauthoryear{Cannon, Cariou  et~al.}{Cannon
  et~al.}{2012}]{early_warning}
Cannon K.,  Cariou R.,   et~al., 2012, \mn@doi [The Astrophysical Journal]
  {10.1088/0004-637x/748/2/136}, 748, 136

\bibitem[\protect\citeauthoryear{Chan, Messenger, Heng  \& Hendry}{Chan
  et~al.}{2018}]{earlywarning_BNS}
Chan M.~L.,  Messenger C.,  Heng I.~S.,   Hendry M.,  2018, \mn@doi [Phys. Rev.
  D] {10.1103/PhysRevD.97.123014}, 97, 123014

\bibitem[\protect\citeauthoryear{Creighton \& Anderson}{Creighton \&
  Anderson}{2011}]{Jolien}
Creighton J. D. E. J. D.~E.,  Anderson W.~G.,  2011, Gravitational-wave physics
  and astronomy : an introduction to theory, experiment and data analysis.
Wiley-VCH, \url {http://ci.nii.ac.jp/ncid/BB07081613}

\bibitem[\protect\citeauthoryear{Cutler \& Flanagan}{Cutler \&
  Flanagan}{1994}]{fisher_CBC}
Cutler C.,  Flanagan E.~E.,  1994, \mn@doi [Phys. Rev. D]
  {10.1103/PhysRevD.49.2658}, 49, 2658

\bibitem[\protect\citeauthoryear{Degallaix~(the
  Virgo~Collaboration)}{Degallaix~(the Virgo~Collaboration)}{2018}]{Virgo_plus}
Degallaix~(the Virgo~Collaboration) J.,  2018, {Advanced Virgo+ preliminary
  studies}, \url{https://tds.virgo-gw.eu/?content=3&r=14287}

\bibitem[\protect\citeauthoryear{{Digel} \& {Myers}}{{Digel} \&
  {Myers}}{2001}]{Fermi_GBM}
{Digel} S.,  {Myers} J.~D.,  2001, {GLAST: Exploring Nature's Highest Energy
  Processes with the Gamma-ray Large Area Space Telescope}, NASA STI/Recon
  Technical Report N

\bibitem[\protect\citeauthoryear{Foucart, Hinderer  \& Nissanke}{Foucart
  et~al.}{2018}]{remnant_mass_function}
Foucart F.,  Hinderer T.,   Nissanke S.,  2018, \mn@doi [Phys. Rev. D]
  {10.1103/PhysRevD.98.081501}, 98, 081501

\bibitem[\protect\citeauthoryear{Harry}{Harry}{2010}]{LIGO2}
Harry G.~M.,  2010, \mn@doi [Classical and Quantum Gravity]
  {10.1088/0264-9381/27/8/084006}, 27, 084006

\bibitem[\protect\citeauthoryear{Husa, Khan, Hannam, P\"urrer, Ohme, Forteza
  \& Boh\'e}{Husa et~al.}{2016}]{PNexample_3.5PN_1}
Husa S.,  Khan S.,  Hannam M.,  P\"urrer M.,  Ohme F.,  Forteza X.~J.,   Boh\'e
  A.,  2016, \mn@doi [Phys. Rev. D] {10.1103/PhysRevD.93.044006}, 93, 044006

\bibitem[\protect\citeauthoryear{Ivezi\'{c} et~al.,}{Ivezi\'{c}
  et~al.}{2019}]{LSST}
Ivezi\'{c} {\v{Z}}.,  et~al., 2019, \mn@doi [The Astrophysical Journal]
  {10.3847/1538-4357/ab042c}, 873, 111

\bibitem[\protect\citeauthoryear{Jefremov, Tsupko  \&
  Bisnovatyi-Kogan}{Jefremov et~al.}{2015}]{ISCO_frequency_KerrBH}
Jefremov P.~I.,  Tsupko O.~Y.,   Bisnovatyi-Kogan G.~S.,  2015, \mn@doi [Phys.
  Rev. D] {10.1103/PhysRevD.91.124030}, 91, 124030

\bibitem[\protect\citeauthoryear{Kapadia, Singh, Shaikh, Chatterjee  \&
  Ajith}{Kapadia et~al.}{2020}]{early_warning_higher_mode}
Kapadia S.~J.,  Singh M.~K.,  Shaikh M.~A.,  Chatterjee D.,   Ajith P.,  2020,
  \mn@doi [The Astrophysical Journal] {10.3847/2041-8213/aba42d}, 898, L39

\bibitem[\protect\citeauthoryear{Khan, Husa, Hannam, Ohme, P\"urrer, Forteza
  \& Boh\'e}{Khan et~al.}{2016}]{PNexample_3.5PN_2}
Khan S.,  Husa S.,  Hannam M.,  Ohme F.,  P\"urrer M.,  Forteza X.~J.,   Boh\'e
  A.,  2016, \mn@doi [Phys. Rev. D] {10.1103/PhysRevD.93.044007}, 93, 044007

\bibitem[\protect\citeauthoryear{Lundgren \& O'Shaughnessy}{Lundgren \&
  O'Shaughnessy}{2014}]{precession1}
Lundgren A.,  O'Shaughnessy R.,  2014, \mn@doi [Phys. Rev. D]
  {10.1103/PhysRevD.89.044021}, 89, 044021

\bibitem[\protect\citeauthoryear{Magee \& Borhanian}{Magee \&
  Borhanian}{2022}]{early_warning_2.5G_dutycycle}
Magee R.,  Borhanian S.,  2022, {Realistic observing scenarios for the next
  decade of early warning detection of binary neutron stars} (\mn@eprint
  {arXiv} {2201.11841})

\bibitem[\protect\citeauthoryear{Magee, Chatterjee  et~al.}{Magee
  et~al.}{2021}]{early_warning_O4}
Magee R.,  Chatterjee D.,   et~al., 2021, {First Demonstration of Early Warning
  Gravitational-wave Alerts}, \mn@doi{10.3847/2041-8213/abed54}, \url
  {https://doi.org/10.3847/2041-8213/abed54}

\bibitem[\protect\citeauthoryear{Maggiore}{Maggiore}{2008}]{maggiore1}
Maggiore M.,  2008, {Gravitational Waves: Volume 1: Theory and Experiments}.
Gravitational Waves, OUP Oxford, \url
  {https://books.google.co.jp/books?id=AqVpQgAACAAJ}

\bibitem[\protect\citeauthoryear{McWilliams \& Levin}{McWilliams \&
  Levin}{2011}]{NSBH_EMemission}
McWilliams S.~T.,  Levin J.,  2011, \mn@doi [The Astrophysical Journal]
  {10.1088/0004-637x/742/2/90}, 742, 90

\bibitem[\protect\citeauthoryear{Metzger}{Metzger}{2019}]{kilonovae}
Metzger B.~D.,  2019, \mn@doi [Living Reviews in Relativity]
  {10.1007/s41114-019-0024-0}, 23

\bibitem[\protect\citeauthoryear{Michimura, Komori  et~al.}{Michimura
  et~al.}{2020}]{KAGRA_plus}
Michimura Y.,  Komori K.,   et~al., 2020, \mn@doi [Phys. Rev. D]
  {10.1103/PhysRevD.102.022008}, 102, 022008

\bibitem[\protect\citeauthoryear{Nakar}{Nakar}{2007}]{prompt_flash}
Nakar E.,  2007, \mn@doi [Physics Reports]
  {https://doi.org/10.1016/j.physrep.2007.02.005}, 442, 166

\bibitem[\protect\citeauthoryear{Nishizawa}{Nishizawa}{2016}]{dNdt_nishizawa}
Nishizawa A.,  2016, \mn@doi [Phys. Rev. D] {10.1103/PhysRevD.93.124036}, 93,
  124036

\bibitem[\protect\citeauthoryear{Nishizawa}{Nishizawa}{2017}]{Measurement_HubbleConstant}
Nishizawa A.,  2017, \mn@doi [Phys. Rev. D] {10.1103/PhysRevD.96.101303}, 96,
  101303

\bibitem[\protect\citeauthoryear{Nitz \& Canton}{Nitz \&
  Canton}{2021}]{early_warning_3G_other_network}
Nitz A.~H.,  Canton T.~D.,  2021, arXiv e-prints, p. arXiv:2106.15259

\bibitem[\protect\citeauthoryear{Nitz, Schäfer  \& Canton}{Nitz
  et~al.}{2020}]{early_warning_2.5G}
Nitz A.~H.,  Schäfer M.,   Canton T.~D.,  2020, \mn@doi [The Astrophysical
  Journal] {10.3847/2041-8213/abbc10}, 902, L29

\bibitem[\protect\citeauthoryear{Punturo et~al.}{Punturo
  et~al.}{2010}]{ET_paper}
Punturo M.,  et~al., 2010, \mn@doi [Classical and Quantum Gravity]
  {10.1088/0264-9381/27/19/194002}, 27, 194002

\bibitem[\protect\citeauthoryear{Sachdev, Magee  et~al.}{Sachdev
  et~al.}{2020}]{early_warning_simulation}
Sachdev S.,  Magee R.,   et~al., 2020, {An Early-warning System for
  Electromagnetic Follow-up of Gravitational-wave Events},
  \mn@doi{10.3847/2041-8213/abc753}, \url
  {https://doi.org/10.3847/2041-8213/abc753}

\bibitem[\protect\citeauthoryear{Schutz}{Schutz}{1986}]{measure_hubble_constant}
Schutz B.,  1986, Nature, 323, 310

\bibitem[\protect\citeauthoryear{Singh, Kapadia, Shaikh, Chatterjee  \&
  Ajith}{Singh et~al.}{2021}]{early_warning_higher_mode_2G2.5G3G}
Singh M.~K.,  Kapadia S.~J.,  Shaikh M.~A.,  Chatterjee D.,   Ajith P.,  2021,
  \mn@doi [Monthly Notices of the Royal Astronomical Society]
  {10.1093/mnras/stab125}, 502, 1612–1622

\bibitem[\protect\citeauthoryear{Smith, Field, Blackburn, Haster, P\"urrer,
  Raymond  \& Schmidt}{Smith et~al.}{2016}]{localization_precession}
Smith R.,  Field S.~E.,  Blackburn K.,  Haster C.-J.,  P\"urrer M.,  Raymond
  V.,   Schmidt P.,  2016, \mn@doi [Phys. Rev. D] {10.1103/PhysRevD.94.044031},
  94, 044031

\bibitem[\protect\citeauthoryear{Somiya}{Somiya}{2012}]{KAGRA1}
Somiya K.,  2012, \mn@doi [Classical and Quantum Gravity]
  {10.1088/0264-9381/29/12/124007}, 29, 124007

\bibitem[\protect\citeauthoryear{Tsang, Read, Hinderer, Piro  \&
  Bondarescu}{Tsang et~al.}{2012}]{resonant_shattering}
Tsang D.,  Read J.~S.,  Hinderer T.,  Piro A.~L.,   Bondarescu R.,  2012,
  \mn@doi [Phys. Rev. Lett.] {10.1103/PhysRevLett.108.011102}, 108, 011102

\bibitem[\protect\citeauthoryear{Tsutsui, Nishizawa  \& Morisaki}{Tsutsui
  et~al.}{2021}]{early_warning_precession}
Tsutsui T.,  Nishizawa A.,   Morisaki S.,  2021, \mn@doi [Physical Review D]
  {10.1103/physrevd.104.064013}, 104

\bibitem[\protect\citeauthoryear{Vecchio}{Vecchio}{2004}]{precession_breaks_degeneracies_LISA}
Vecchio A.,  2004, \mn@doi [Physical Review D] {10.1103/physrevd.70.042001}, 70

\bibitem[\protect\citeauthoryear{Vitale, Lynch, Veitch, Raymond  \&
  Sturani}{Vitale et~al.}{2014}]{precession_breaks_degeneracies_ground}
Vitale S.,  Lynch R.,  Veitch J.,  Raymond V.,   Sturani R.,  2014, \mn@doi
  [Physical Review Letters] {10.1103/physrevlett.112.251101}, 112

\bibitem[\protect\citeauthoryear{Wei \& Huerta}{Wei \&
  Huerta}{2021}]{early_warning_neural_network3}
Wei W.,  Huerta E.,  2021, \mn@doi [Physics Letters B]
  {https://doi.org/10.1016/j.physletb.2021.136185}, 816, 136185

\bibitem[\protect\citeauthoryear{Will \& Wiseman}{Will \&
  Wiseman}{1996}]{PNexample_1.5PN}
Will C.~M.,  Wiseman A.~G.,  1996, \mn@doi [Physical Review D]
  {10.1103/physrevd.54.4813}, 54, 4813–4848

\bibitem[\protect\citeauthoryear{{Yu}, {Adhikari}, {Magee}, {Sachdev}  \&
  {Chen}}{{Yu} et~al.}{2021}]{early_warning_neural_network1}
{Yu} H.,  {Adhikari} R.~X.,  {Magee} R.,  {Sachdev} S.,   {Chen} Y.,  2021,
  arXiv e-prints, \href {https://ui.adsabs.harvard.edu/abs/2021arXiv210409438Y}
  {p. arXiv:2104.09438}

\makeatother
\end{thebibliography}

% Alternatively you could enter them by hand, like this:
% This method is tedious and prone to error if you have lots of references
%\begin{thebibliography}{99}
%\bibitem[\protect\citeauthoryear{Author}{2012}]{Author2012}
%Author A.~N., 2013, Journal of Improbable Astronomy, 1, 1
%\bibitem[\protect\citeauthoryear{Others}{2013}]{Others2013}
%Others S., 2012, Journal of Interesting Stuff, 17, 198
%\end{thebibliography}

%%%%%%%%%%%%%%%%%%%%%%%%%%%%%%%%%%%%%%%%%%%%%%%%%%

%%%%%%%%%%%%%%%%% APPENDICES %%%%%%%%%%%%%%%%%%%%%

%\appendix
%
%\section{Some extra material}
%
%If you want to present additional material which would interrupt the flow of the main paper,
%it can be placed in an Appendix which appears after the list of references.

%%%%%%%%%%%%%%%%%%%%%%%%%%%%%%%%%%%%%%%%%%%%%%%%%%

% Don't change these lines
\bsp	% typesetting comment
\label{lastpage}
\end{document}